\documentclass[aip,apl,preprint]{revtex4-1}
\usepackage{graphicx}
\usepackage{bm}
\usepackage{epstopdf}
\usepackage{amssymb}
\usepackage{amsmath}
\usepackage{natbib}
\usepackage{mathcomp}

\begin{document}

\title{A ballistic $pn$ junction in suspended graphene with split bottom gates}

\author{Anya L. Grushina}
\author{Dong-Keun Ki}
\author{Alberto F. Morpurgo}
\email[]{Alberto.Morpurgo@unige.ch}
\affiliation{D\'{e}partment de Physique de la Mati\'{e}re
Condens\'{e}e (DPMC) and Group of Applied Physics (GAP), University
of Geneva, 24 Quai Ernest-Ansermet, CH1205 Gen\'{e}ve,
Switzerland}

\date{\today}

\begin{abstract}
We have developed a process to fabricate suspended
graphene devices with local bottom gates, and tested it by realizing electrostatically 
controlled  $pn$ junctions on a suspended graphene mono-layer nearly 2 $\mu$m long. 
Measurements as a function of gate voltage, magnetic field, bias, and temperature exhibit characteristic
Fabry-Perot oscillations in the cavities formed by the $pn$ junction and each
of the contacts, with transport occurring in the ballistic regime. 
Our results demonstrate the possibility to achieve  a high degree of control on the local electronic properties of
ultra-clean suspended graphene layers, a key aspect for the realization of new graphene nanostructures.
\end{abstract}


\maketitle 

Technical developments in device fabrication are essential to
perform transport experiments revealing the intrinsic electronic properties of graphene.
Suspended graphene devices~\cite{Bolotin2008,Du2008} and devices with hexagonal
boron nitride (hBN) as supporting substrate~\cite{Dean2010,Dean2012}
provide clear examples. The same is true for double-gated devices,
in which graphene is not in direct contact with any dielectric
material.~\cite{Weitz2010,Velasco2012,Allen2012} Indeed, these
devices have allowed the observation of phenomena such as the
fractional quantum Hall effect,~\cite{Du2009,Bolotin2009,Dean2011,Ki2013a} new
interaction-induced symmetry broken states in
bilayers,~\cite{Weitz2010,Velasco2012} and manifestations of
ballistic transport.~\cite{Mayorov2011,Campos2012,Ki2013} Even more advanced experiments
would be possible if double-gating on suspended devices could be performed locally.
In bilayer graphene, for instance, local double gating would allow the study of topological confinement,~\cite{Martin2008}
and the realization of fully electrostatically tunable $pn$ junctions, of
interest to generate or detect light at continuously tunable
frequencies in the THz to mid infrared range. As an essential step
towards the realization of these new structures, here we describe a
technique to fabricate high-quality suspended graphene devices with local
bottom gates and apply it to the realization of an electrostatically tunable, 
ballistic $pn$ junction in monolayer graphene.

\begin{figure}[t!]
\includegraphics[width=8.5cm]{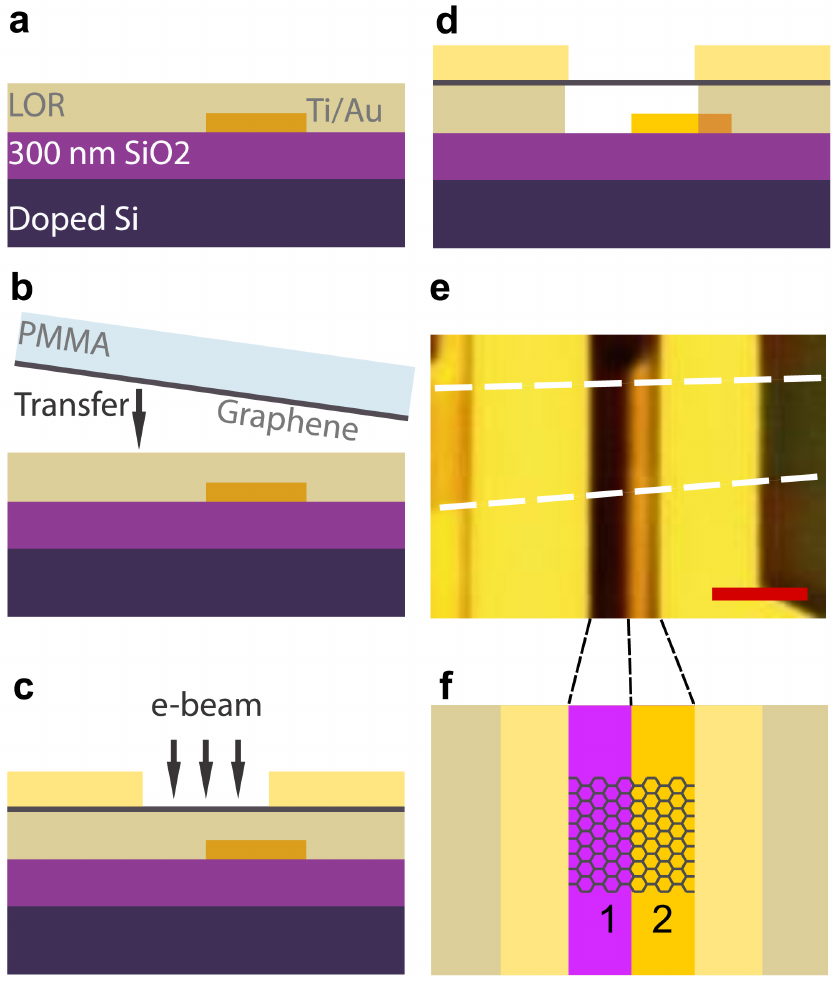}
\caption{(Color online) Schematic summary of the fabrication process (drawings not
to scale). (a) A target substrate with predefined gate electrodes (10 nm Ti/30 nm Au) is covered 
with a 450-nm-thick LOR layer. (b) A graphene flake on a PMMA support is
transferred onto LOR, aligned to the bottom electrode. (c-d) Graphene is contacted 
and the underlying LOR exposed with an electron beam, to achieve suspension. 
(e) Optical microscope image of the device (the dashed lines indicate the edges of
the graphene flake); the local bottom gate is visible under the right
electrode (the  bar is 2 $\mu$m long). (e) Schematic top view
denoting the regions 1 and 2 (coupled primarily to the two different gate electrodes).}
\label{Fig.1}
\end{figure}

The fabrication process is illustrated schematically in Fig.
\ref{Fig.1}. The first step consists in preparing the bottom gates
in the desired configuration --in the present case, a simple single
strip-- on a doped silicon substrate covered with 300 nm SiO$_2$, by
using conventional techniques (electron-beam lithography, Ti/Au
evaporation, and lift-off). Next, a 450-nm-thick layer of
polydimethylglutarimide (PMGI)-based lift-off resist (LOR,
MicroChem) is spun onto the substrate (Fig. \ref{Fig.1}(a)).
LOR resist is chosen because it is not only compatible with all
subsequent micro-fabrication processes, but also it can be exposed with an
electron beam and developed away to suspend graphene at the end of
the fabrication process.~\cite{Tombros2011,Tombros2011a,Ki2012}

As a second step, a graphene flake is transferred onto the LOR layer,
and positioned on to the bottom gate (Fig. \ref{Fig.1}(b)).
To this end, we adapted a technique developed to fabricate
graphene/hBN heterostructures.~\cite{Dean2010} Specifically,
graphene is exfoliated using an adhesive tape and placed on a different
substrate, previously coated with a layer of water soluble polymer
(a 9 wt.$\%$ poly(4-styrenesulfonic acid) solution in water) and a
layer of PMMA.~\cite{Nam2011} After the desired flake is identified under 
an optical microscope, the substrate is immersed in
water, causing the water-soluble polymer to dissolve and the PMMA to
float. The floating PMMA is retrieved using a plastic support, which
is then mounted onto a micro-manipulator under an optical microscope.
This enables the graphene flake to be transferred onto the LOR, aligned to the bottom gate with
a precision of a few microns (Fig. \ref{Fig.1}(b)). After securing it by heating at 105 $\tccelsius$
for 40 minutes, the flake is contacted with Ti/Au electrodes
(10/60 nm thick) defined by conventional electron-beam lithography,
metal evaporation, and lift-off (for PMMA on LOR, development and
lift-off are done using Xylene,~\cite{Tombros2011,Tombros2011a,Ki2012}
at room temperature and $\sim$90 $\tccelsius$, respectively). In
the final step, the LOR under the graphene layer is exposed with an
electron beam (Fig. \ref{Fig.1}(c)) and developed away to achieve
the suspension (Fig. \ref{Fig.1}(d)).~\cite{Tombros2011,Tombros2011a,Ki2012}

\begin{figure}[t!]
\includegraphics[width=8.5cm]{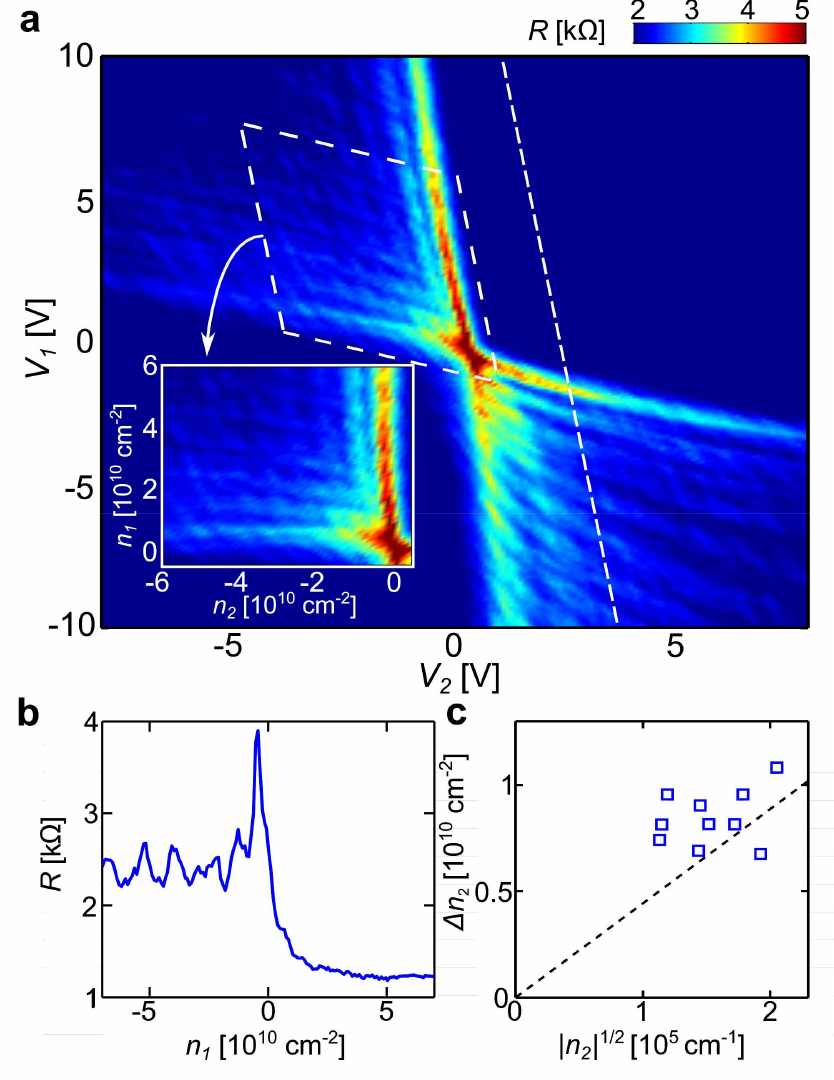}
\caption{(Color online) (a) Resistance oscillation measured at $T=0.25$
K as a function of $V_1$ and $V_2$. Along the dashed white line,
$n_1$ changes while $n_2$ is kept fixed at $2.82\times10^{10}$ cm$^{-2}$. 
The data shown in a panel (b), and in Figs. \ref{Fig.3} and \ref{Fig.4}
are taken along this line. The inset shows the plot of $R$ as a function
of $n_1$ and $n_2$, for the $V_1$ and $V_2$ region corresponding to the parallelogram delimited by the dashed line. 
(b) $R(n_1)$ at fixed, positive $n_2$ shows oscillations for negative $n_1$, i.e., 
when a $pn$ junction is present. (c) Oscillation period $\Delta n_2$, plotted as a function of
$\sqrt{|n_2|}$. Empty squares show data taken at different values of $n_1$;
the broken line represents the values of $\Delta n_2 =2\sqrt{\pi|n_2|}/L_2$ estimated form a simple
particle-in-a-box approximation ($L_2\approx800$ nm).} \label{Fig.2}
\end{figure}

We have applied this technique to suspend a 1.8 $\mu$m long graphene monolayer over a
bottom gate that overlaps with about half of the suspended length (see Fig. \ref{Fig.1}(e)). 
By screening the potential generated by a voltage applied to the conducting Si
substrate, the bottom gate defines two regions (1 and 2, see Fig. \ref{Fig.1}(f)),
whose carrier density and type can be controlled by applying
voltages to the doped silicon substrate ($V_1$) and to the local gate
itself ($V_2$). Transport measurements as a function of $V_1$ and
$V_2$ were performed in a Heliox He$^3$ system to characterize the
device at different magnetic field ($B$), bias ($V_{DC}$), and
temperature ($T$). Prior to the measurements, the device was
annealed at 4.2 K by passing a sufficiently large current through
the graphene flake.

Fig. \ref{Fig.2}(a) shows the resistance $R$ measured at $T=0.25$ K
as a function of $V_1$ and $V_2$. Four quadrants can be identified,
roughly corresponding to $V_1$ and $V_2$ having the same or opposite
sign. When the sign is the same, no $pn$ junction is present in the
device: only either electrons or holes are accumulated in regions 1
and 2. A $pn$ junction is present between region 1 and
2 when  $V_1$ and $V_2$ have opposite sign. The borders
of the different quadrants are not parallel to the $V_1$ and $V_2$
axis, because of the cross-talk between two gates: $V_1$ does not
only change the density in  region 1 ($n_1$) but also --to a
lesser extent-- the density in region 2 ($n_2$); similarly, $V_2$
also influences the density $n_1$. Although, in general, that
the density is not spatially uniform in regions 1 and 2 (this is obvious when a $pn$ junction is present,
in which case the carrier density vanishes at the interface between the two regions), accounting as much as
possible for the effect of the cross-talk is useful to analyze the
data.  This can be done by looking at the gate and magnetic field
dependence of the quantized Hall conductance plateaus in the unipolar
regime, where $n_1\simeq n_2$ (i.e., when the density non uniformity is
less pronounced). We find $n_1$ [10$^{10}$ cm$^{-2}$] $=1.0\times
V_1$ [V] + $0.35\times V_2$ [V] + 0.5 and $n_2$ [10$^{10}$
cm$^{-2}$] $=0.2\times V_1$ [V] + $1.4\times V_2$ [V] $-$ 0.4 (the
constants account for the shift of charge neutrality point from
$V_{1/2}=0$ V; the proportionality terms between $n_{1/2}$ and
$V_{1/2}$ are in good agreement with the estimated geometrical capacitances). The resistance as a
function of $n_1$ and $n_2$ defined in this way is shown in the
inset of Fig. \ref{Fig.2}(a).

\begin{figure}[t!]
\includegraphics[width=8.5cm]{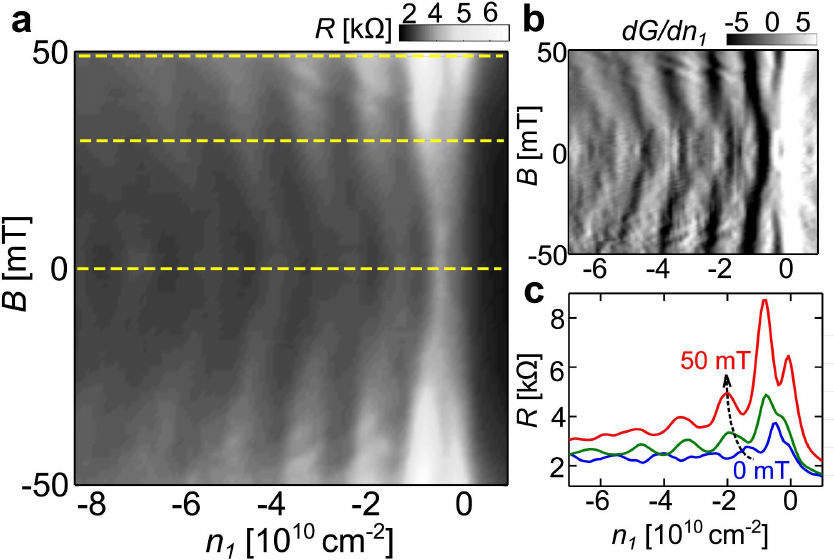}
\caption{(Color online) Magnetic field dependence of $R(n_1)$
measured at $T=0.25$ K, at fixed $n_2$, along the dashed line shown in Fig.
\ref{Fig.2}(a). (a) Plot of $R(n_1,B)$, showing characteristic $\pi$-shift at $B=B^*\approx20$-30 mT. 
(b) Plot of $dG/dn_1$, as a function of $n_1$ and $B$, shown for comparison with similar data reported in 
the literature. In (a), the broken lines indicate the values of $B$ ($B=0$, 30, 50 mT ) at which the data 
in (c) are measured.  } \label{Fig.3}
\end{figure}

\begin{figure}[t!]
\includegraphics[width=8.5cm]{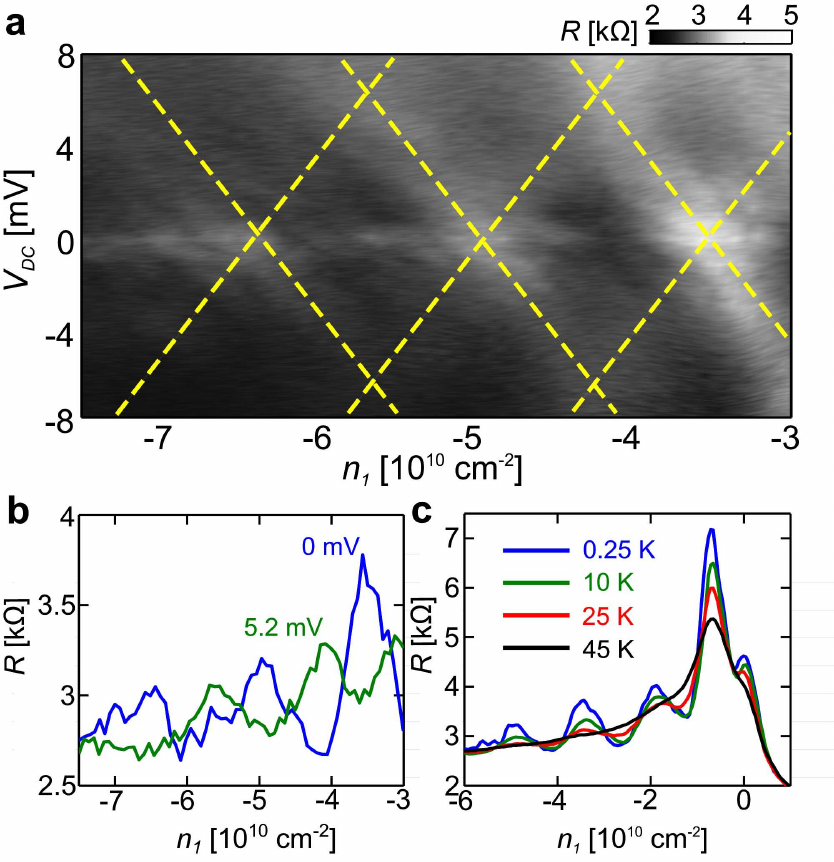}
\caption{(Color online) Energy dependence of the Fabry-Perot interference.
(a) Resistance $R$ measured at
$T=0.25$ K as a function of bias $V_{DC}$ and $n_1$ (at fixed $n_2$ along a dashed line in
Fig. \ref{Fig.2}(a)); the yellow dashed lines are guides to the eye . 
(b) Two representative curves
of $R(n_1)$ measured at at $V_{DC}=0$ and 5.2 mV, showing a $\pi$ shift in the oscillation phase. 
(c) $T$-dependence of $R(n_1)$ measured at $V_{DC}=0$ mV. All data in this figure were taken at $B=40$ mT} \label{Fig.4}
\end{figure}

When $V_1$ and $V_2$ are biased with opposite polarity to create a $pn$ junction, the resistance doubles as compared to when no $pn$ junction is present (compare, e.g., the
resistance for $n_1<0$ and $n_1>0$ in Fig. \ref{Fig.2}(b)). This shows that the $pn$ junction
gives a large contribution to the total device resistance, despite the sizable device length (1.8 $\mu$m in total).
In particular, the $pn$ junction contribution significantly larger as
compared to previously studied $pn$ and $pnp$ junctions on SiO$_2$
substrates.~\cite{Huard2007,Williams2007a,Gorbachev2008,Stander2009}

Fig. \ref{Fig.2}(a) further shows that the resistance also oscillates as a function of $V_1$ and 
$V_2$ when a $pn$ junction is formed, in a way resembling the behavior of graphene $pnp$ junctions on a Si/SiO$_2$
substrate.~\cite{Young2009,Nam2011} In that case, the oscillations were shown to originate from Fabry-Perot interference
of Dirac electrons moving ballistically within the ($\approx 100$ nm
long) cavity defined by the $pnp$ region.~\cite{Young2009,Nam2011,Shytov2008} 
In our device, Fabry-Perot oscillations occur in
cavities formed by the $pn$ junction and each of the two interfaces with the metal contacts, where carriers
are also backscattered.~\cite{Miao2007,Cho2011} Our device therefore consists
of two Fabry-Perot cavities connected in series, and the "checkerboard" pattern visible in Fig. \ref{Fig.2}(a) is a
manifestation of interference in both cavities. The clear visibility of the oscillations directly in the resistance 
(Figs. \ref{Fig.2}(a) and \ref{Fig.2}(b)), without the need of derivating the data, is indicative of the high quality of 
the suspended $pn$ junction.~\cite{Young2009,Nam2011}

An estimate of the oscillation period $\Delta n$ (i.e., the distance in density between two nearest resistance peaks or dips) is obtained by imposing that the dynamical phase acquired by an electron wave propagating back and forth in the cavity is equal to $2\pi$, i.e. $\Delta(2k_{F,i}L_i)=2\pi$ (the subscript $i=1,2$ label the region). As $k_{F,i}=\sqrt{\pi n_i}$, we obtain $\Delta n_i=2\sqrt{\pi n_i}/L_i$ (note that several previous references~\cite{Young2009,Nam2011,Velasco2009} reported an
incorrect expression, $\Delta n_i=4\sqrt{\pi n_i}/L_i$, differing by a factor of 2 from ours).
The dotted line in Fig. \ref{Fig.2}(c) represents the values of $\Delta n_2$ estimated using this formula
for region 2 ($L_2=800$ nm), and the open squares are the experimental values extracted from the most
pronounced oscillations measured upon changing $n_2$ (a similar result is obtained for region 1). The order of 
magnitude and the trend in the data are well captured by the simple theoretical expression. The experimental values, however, are somewhat larger than expected, because the carrier density in the
region close to the $pn$ junction is lower than the calculated value
$n_2$. The lower density causes a smaller value of $k_F$, and therefore a smaller phase shift 
and an additional increase in carrier density is needed to compensate for this effect.

The evolution of the oscillation phase upon increasing magnetic field $B$ (see Fig. \ref{Fig.3}) provides further evidence
for the Fabry-Perot nature of the interference.~\cite{Shytov2008,Young2009,Nam2011} Fig. \ref{Fig.3}(a)
shows the $B$-dependence of the oscillations upon changing $n_1$ at
fixed $n_2=2.82\times10^{10}$ cm$^{-2}$ (i.e., by changing $V_1$ and $V_2$ along the dashed line
depicted in Fig. \ref{Fig.2}(a)), which exhibits a $\pi$ phase shift at $B\equiv B^*\approx20$-30 mT (varying $n_2$ at
fixed $n_1$ gives comparable results). Fig. \ref{Fig.3}(b) shows the same effect in the derivative of the conductance ($G = 1/R$) with respect to $n_1$, and panel (c) illustrates the occurrence of the phase shift, with three individual slices of the color plot shown in (a), taken at $B=0,$ 30, and 50 mT.

As discussed for $pnp$ junctions,~\cite{Shytov2008,Nam2011,Young2009} the phase shift
originates from the unique properties of Dirac electrons, namely the
angular dependence of the reflection probability at a $pn$
junction,~\cite{Katsnelson2006,Cheianov2006} and the accumulation of
a $\pi$ Berry phase along momentum-space trajectories 
that enclose the origin.~\cite{Novoselov2005,Zhang2005} For a given position of the Fermi energy, the
electrons contributing predominantly to the Fabry-Perot resistance
oscillations are those incident on the $pn$ junction with a certain
transverse momentum ($k_{y0}$; the specific value
depends on the density profile across the
junction).~\cite{Shytov2008,Young2009} Upon increasing the
perpendicular magnetic field, the electron trajectories in the
Fabry-Perot cavity are bent, and --in momentum space-- they
eventually enclose the origin.~\cite{Young2009,Shytov2008} When this
happens, an additional Berry phase $\pi$ is acquired, causing the phase shift in the resistance oscillations. For 
$\simeq 100$ nm long $pnp$ junctions on substrate, the shift was found to occur at
$B^*\approx2\hbar k_{y0}/eL\approx250$-500
mT.~\cite{Young2009,Nam2011} Assuming a comparable value
of $k_{y0}$ (within a factor of 2-3), this is consistent with 
our observations: the phase shift occurs at an order of magnitude smaller $B^*\simeq20$-30
mT, corresponding to an order of magnitude longer cavity.

Finally, we discuss the characteristic energy scale of the
resistance oscillations. Fig. \ref{Fig.4}(a) shows the differential
resistance measured as a function of bias $V_{DC}$ and density
$n_1$. Systematically, the position of the resistance peaks shifts
linearly upon increasing $V_{DC}$, as expected for Fabry-Perot
interference.~\cite{Miao2007,Cho2011} The shift is also illustrated
by Fig. \ref{Fig.4}(b), which compares measurements taken at
$V_{DC}=0$ and $5.2$ mV. From both Figs. \ref{Fig.4}(a) and
\ref{Fig.4}(b), the bias needed to shift a maximum of differential resistance
into a minimum is approximately 5 meV. We have also looked at the energy dependence of the
oscillation by changing temperature, and found that the oscillations
are washed out at about 40 K ($\approx3.5$ meV). Since, owing to the non-uniform charge density, the level spacing
in the cavity is somewhat larger than the particle-in-a-box value $hv_F/2L\approx2.5$ meV (with
$L \simeq 1 \mu$m and $v_F=10^6$ m/s), the energy scale found in the experiments is consistent with the 
simplest theoretical estimate.

We conclude that the behavior of our device is consistent
with the presence of a $pn$ junction, and with transport occurring in the
ballistic regime over a length comparable to the device size (1.8 $\mu$m). 
The measurements therefore confirm that the fabrication technique that enables the realization of suspended
graphene devices with local bottom gates preserves the high quality of the material.
In the future, this technology will be applied to realize new graphene devices relying on 
the local control of the electrostatic potential and electric field, such as the nano-structures
needed for the study of Veselago lensing,~\cite{Cheianov2007}
of collimation of electrons,~\cite{Park2008} and of topological
confinement.~\cite{Martin2008}

We gratefully acknowledge A. Ferreira for technical support and
financial support from the SNF, NCCR MaNEP, and NCCR QSIT.

\end{document}